\documentclass[lettersize,journal]{IEEEtran}
\usepackage{algorithmic}
\usepackage{array}
\usepackage[caption=false]{subfig}
\usepackage{textcomp}
\usepackage{stfloats}
\usepackage{url}
\usepackage{verbatim}
\usepackage{graphicx}
\usepackage{cite}
\usepackage{amsthm}
\usepackage{siunitx}
\usepackage{amsmath,amssymb,amsfonts}
\usepackage{xcolor}
\usepackage{overpic} 
\usepackage{cuted}
\usepackage[linesnumbered,ruled,noend]{algorithm2e}
\usepackage{bm}
\usepackage{booktabs}
\begin{document}

\title{STAR‑RIS-assisted Collaborative Beamforming for Low-altitude Wireless Networks}

\author{
Xinyue Liang, 
Hui Kang, 
Junwei Che, 
Jiahui Li,~\IEEEmembership{Member,~IEEE,} 
Geng Sun,~\IEEEmembership{Senior Member,~IEEE}, \\
Qingqing Wu,~\IEEEmembership{Senior Member,~IEEE},  
Jiacheng Wang, 
and Dusit Niyato,~\IEEEmembership{Fellow,~IEEE}

\thanks{

        \par Xinyue Liang, Hui Kang, Junwei Che, and Jiahui Li are with the College of Computer Science and Technology, Jilin University, Changchun 130012, China (e-mails: xyliang25@mails.jlu.edu.cn; kanghui@jlu.edu.cn; jwche24@163.com; lijiahui@jlu.edu.cn). 

        \par Geng Sun is with the College of Computer Science and Technology, Jilin University, Changchun 130012, China, and with Key Laboratory of Symbolic Computation and Knowledge Engineering of Ministry of Education, Jilin University, Changchun 130012, China; he is also affiliated with the College of Computing and Data Science, Nanyang Technological University, Singapore 639798 (e-mail: sungeng@jlu.edu.cn).
        
        \par Qingqing Wu is with the Department of Electronic Engineering, Shanghai Jiao Tong University, Shanghai 200240, China (e-mail: qingqingwu@sjtu.edu.cn).

        \par Jiacheng Wang and Dusit Niyato are with the College of Computing and Data Science, Nanyang Technological University, Singapore 639798 (e-mails: jiacheng.wang@ntu.edu.sg; dniyato@ntu.edu.sg).
        \par \textit{Corresponding authors: Geng Sun; Jiahui Li.} 
        \protect
	}
}

\maketitle
\begin{abstract}

While low-altitude wireless networks (LAWNs) based on uncrewed aerial vehicles (UAVs) offer high mobility, flexibility, and coverage for urban communications, they face severe signal attenuation in dense environments due to obstructions. To address this critical issue, we consider introducing collaborative beamforming (CB) of UAVs and omnidirectional reconfigurable beamforming (ORB) of simultaneous transmitting and reflecting reconfigurable intelligent surfaces (STAR-RIS) to enhance the signal quality and directionality. 
On this basis, we formulate a joint rate and energy optimization problem (JREOP) to maximize the transmission rate of the overall system, while minimizing the energy consumption of the UAV swarm. Due to the non-convex and NP-hard nature of JREOP, we propose a heterogeneous multi-agent collaborative dynamic (HMCD) optimization framework, which has two core components. The first component is a simulated annealing (SA)-based STAR-RIS control method, which dynamically optimizes reflection and transmission coefficients to enhance signal propagation. The second component is an improved multi-agent deep reinforcement learning (MADRL) control method, which incorporates a self-attention evaluation mechanism to capture interactions between UAVs and an adaptive velocity transition mechanism to enhance training stability. Simulation results demonstrate that HMCD outperforms various baselines in terms of convergence speed, average transmission rate, and energy consumption. Further analysis reveals that the average transmission rate of the overall system scales positively with both UAV count and STAR-RIS element numbers.

\end{abstract}
	

%

\section{Introduction}\label{sec:introduction}

\par Low-altitude wireless networks (LAWNs), which aim to extend the capabilities of terrestrial networks through low-altitude aerial platforms, have emerged as an important paradigm for sixth-generation (6G) wireless communications. Among various enabling platforms for LAWNs, uncrewed aerial vehicles (UAVs) play a central role since they provide high mobility, flexible deployment and broad coverage. Specifically, due to the properties of high line-of-sight (LoS) probability, UAVs can be deployed in remote areas as flying base stations to achieve low-cost and flexible Internet access and network coverage~\cite{Zhang2024}. Meanwhile, in the scenarios where ground base stations are destroyed by natural disasters or unexpected events, UAVs can be rapidly deployed to establish emergency communication systems~\cite{Sun2025}. Beyond communication coverage, UAVs equipped with the onboard computing resources can collect and process data from Internet-of-Things (IoT) devices that are deployed in complex terrains, thereby offering flexible computational support for wireless applications~\cite{AlBakhrani2025}. Despite these advantages, UAV-based LAWNs still face significant challenges in urban environments, as buildings and other obstacles frequently block propagation paths, thus leading to signal attenuation and transmission interruptions. These issues highlight the need for effective approaches to enhance signal strength and regulate signal direction. 

\par Among potential solutions for enhancing the signal strength, collaborative beamforming (CB) is a promising approach because it enables the distributed UAVs to jointly transmit signals without altering hardware configurations~\cite{Du2025}. Specifically, by forming UAV virtual antenna array (UVAA), multiple UAVs can steer high-gain beams toward legitimate users while minimizing signal leakage to unintended directions, thus achieving both long-range and secure wireless communication. This coordinated transmission mechanism significantly enhances the spectral efficiency and energy utilization. 

\par Although CB can enhance signal strength, direct signal paths between UAVs and ground terminals remain vulnerable to complete blockage by large urban structures~\cite{Zhang2024a}. To address this limitation, LAWNs can leverage reconfigurable intelligent surfaces (RIS) that are deployed on building facades or other infrastructure to bypass obstacles. In particular, compared with traditional reflection-only RIS which is limited to half-space coverage, simultaneous transmitting and reflecting RIS (STAR-RIS) provides a more versatile full-space solution~\cite{Su2023}. Specifically, STAR‑RIS can redirect incident signals to users on both sides of the surface, thus enabling omnidirectional reconfigurable beamforming (ORB) with \ang{360} signal coverage and significantly enhancing the flexibility of system deployment. Consequently, the joint use of UVAA and STAR-RIS enables a hybrid active–passive beamforming framework that holds strong potential for energy-efficient LAWNs.

\par While the joint CB and ORB system holds great promise, its practical realization faces several significant technical hurdles. Firstly, the system exhibits inherent dynamicity, as user positions change over time and real-time coordination between UAVs and STAR-RIS must adapt continuously to evolving environmental conditions. Secondly, the system presents dual-objective conflicts between maximizing the transmission rate of the overall system and minimizing the energy consumption of the UAV swarm. Finally, the high-dimensional complexity arising from heterogeneous devices creates a vast optimization space with non-convex characteristics, where conventional algorithms face significant scalability challenges. These three properties, namely dynamicity, dual-objective trade-offs, and high-dimensional complexity, necessitate the development of scalable optimization methods, which are capable of handling time-varying scenarios, balancing competing objectives, and addressing non-convex optimization challenges. Based on the above analysis, we propose a unified optimization framework for STAR‑RIS‑assisted UVAA. The main contributions of our work are as follows:

\begin{itemize}
    \item \textit{Flexible and Efficient Low-Altitude Communication System:} We design a novel flexible communication architecture that integrates UVAA and STAR-RIS, which can provide \ang{360} signal coverage capability. Specifically, our approach dynamically optimizes the CB of the UVAA while concurrently configuring the dual-functional reflection and transmission properties of the STAR-RIS. Our core objective is to enhance energy efficiency while enabling reliable data transmission in the scenarios with complex obstacles and mobile users. To the best of our knowledge, this is the first framework that leverages both the mobility of UVAA-based CB and full-space coverage capability of STAR-RIS to achieve truly flexible omnidirectional communication.
    
    \item \textit{Dynamic Non-convex NP-hard Optimization Problem:} Within our flexible low-altitude communication system, we formulate a joint rate and energy optimization problem (JREOP). This problem aims to maximize the transmission rate of the overall system while minimizing the energy consumption of the UAV swarm, by optimizing the excitation current weights and flight trajectories of UVAA, as well as the transmission and reflection matrices of STAR-RIS. We demonstrate that JREOP is a non-convex and NP-hard optimization problem that exhibits distinct dynamic and heterogeneous characteristics. 
    
    \item \textit{Heterogeneous Decision Optimization Framework:} Given the dynamic and heterogeneous nature of JREOP, we transform it into a heterogeneous Markov decision process (MDP) and propose the heterogeneous multi-agent collaborative dynamic (HMCD) optimization framework. This framework has two core components, namely an adaptive temperature-based STAR-RIS optimization (ATSO) strategy for STAR‑RIS control and a multi‑agent deep reinforcement learning (MADRL) method for UVAA coordination. Specifically, the UVAA control strategy incorporates two key improvements, including a self-attention evaluation mechanism that captures the inter-agent interactions and an adaptive velocity transition mechanism that enhances the stability of HMCD.
    
    \item \textit{Simulations and Analyses:} Simulation results demonstrate that the proposed HMCD effectively solves the JREOP and significantly outperforms various baselines, such as multi-agent soft actor-critic (MASAC) and multi-agent deep deterministic policy gradient (MADDPG), in terms of convergence speed, average transmission rate, and energy consumption. Moreover, further analysis reveals that the transmission rate scales positively with both the number of UAVs and STAR-RIS elements. 
\end{itemize}    

\par The rest of this paper is organized as follows. Section \ref{sec:related_works} reviews the related research activities. Section \ref{sec:system_model} presents the system model and preliminary explanations. Section \ref{sec:formulation} formulates the optimization problem. Section \ref{sec:algorithm} proposes the optimization framework. Section \ref{sec:simulation} demonstrates the simulation results, and Section \ref{sec:conclusion} concludes the paper.

%
\section{Related Work}\label{sec:related_works}
         
\par This section briefly reviews related work on intelligent device-assisted communication paradigms, joint rate–energy optimization frameworks, and UAV-supported optimization methods to emphasize our contributions.

\subsection{Communication Paradigms for 6G LAWNs}

\par The introduction of more flexible and energy-efficient devices, particularly UAVs and STAR-RIS, into 6G LAWNs has created new opportunities to enhance transmission performance by strengthening signal power and redirecting signal propagation directions. For example, the authors in~\cite{Li2024} investigated a scheme that applies CB to both IoT and UAV systems, which enables efficient data collection and transmission from multiple IoT clusters to a remote base station, along with improvements in energy efficiency and latency. 
Apart from UAVs, STAR-RIS offers a new solution for controlling wireless signal propagation by precisely adjusting the amplitude and phase shift of its transmission and reflection elements. As an illustration, the authors in~\cite{Guo2023} maximized the energy efficiency of the system in time-varying channels by dynamically and jointly optimizing the beamforming vectors at the base station and the STAR-RIS coefficient matrix. 
\par However, despite these advancements in using UAVs and STAR-RIS separately in 6G LAWNs, existing studies have not yet considered the joint application of UAV-based CB and STAR-RIS. In particular, UAV-based CB may be less effective in the presence of obstacles, while most STAR-RIS studies assume static base stations and users, thus overlooking the network dynamics.

\subsection{Optimization Frameworks for Joint Rate-Energy Efficiency}

\par Security strategies that jointly consider UAVs and STAR-RIS primarily focus on optimizing the trajectory of UAVs, transmit power of UAVs, and CB of multi-antenna UAVs, with the assistance of STAR-RIS. For instance, the authors in~\cite{Wang2025} ensured communication quality for distant users and communication concealment for nearby users in UAV-based non-orthogonal multiple access (NOMA) network, by jointly optimizing the CB and ORB of a single UAV and STAR-RIS, as well as the flight trajectory of the UAV. 
Furthermore, the authors in~\cite{Zhao2024} aimed to maximize the minimum average secrecy rate through joint optimization of the beamforming of a multi-antenna UAV base station, UAV trajectory, and transmission and reflection coefficients of the STAR-RIS. 

\par However, despite these advances in security-focused optimization for UAV and STAR-RIS systems, there remains a significant research gap regarding the joint optimization of transmission rate and energy efficiency, particularly in STAR-RIS and multi-UAV cooperative systems.

\subsection{Optimization Methods for UAV-Supported Systems}

\par In 6G mobile networks, the combination of CB (through UAVs) and ORB (through STAR-RIS) represents a forward-looking research direction. Existing studies on related algorithms are generally divided into three categories, which are iterative algorithms, heuristic algorithms, and reinforcement learning algorithms. For example, the authors in~\cite{Su2023} formulated a sum rate maximization problem in a STAR-RIS-assisted NOMA-UAV network, where they transformed the non-convex problem into a convex one and then proposed an efficient iterative algorithm to obtain a suboptimal solution. 
Furthermore, the authors in~\cite{Zhao2022} proposed a sum rate maximization problem for UAV system assisted by STAR-RIS, and introduced a reinforcement learning algorithm based on distributed robust optimization to handle the uncertainty in the locations of obstacles in the environment. 

\par However, algorithms that simultaneously consider the heterogeneity of STAR-RIS and UAVs are still limited. The high degree of system dynamics and heterogeneity introduces a large number of coupled decision variables, thus making traditional iterative or heuristic algorithms difficult to converge efficiently. 

\subsection{Research Positioning and Contributions}

\par Different from existing studies, our work aims to design a unified optimization framework for STAR‑RIS‑assisted UVAA, with joint consideration of both system transmission rate and energy efficiency. To this end, we jointly optimize the UVAA-based CB, STAR-RIS-based ORB, and flight trajectories of the UAV swarm, while employing an MADRL-based policy training approach to tackle the complex control problem. In what follows, we present the system model and relevant preliminaries, thereby characterizing the relationships between the decision variables and the transmission performance and energy efficiency of the system.

%
\section{System Model}\label{sec:system_model}

%
\begin{figure}
    \centering
    \includegraphics[width=1.0\linewidth]{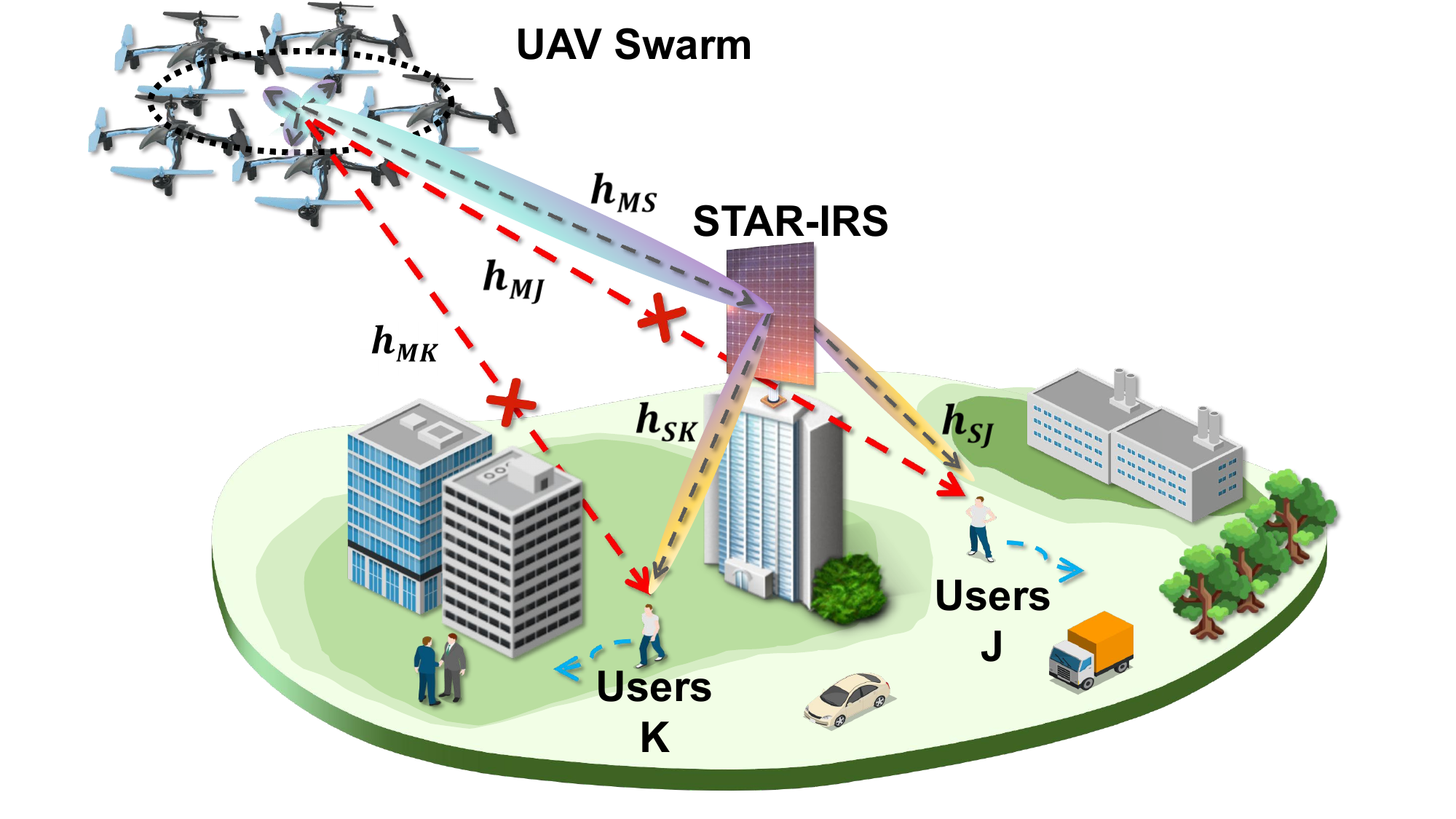}
    \caption{The model of the STAR-RIS-assisted UVAA communication system using joint CB and ORB while the direct channel is obscured.}
    \label{Fig:system_model}
\end{figure}

\par In this section, we first present an overview of the STAR-RIS-assisted UVAA communication system. Then, we detail the transmission model and UAV energy cost model of the system to obtain the key decision variables. 

%
\subsection{System Overview}

\par As shown in Fig. \ref{Fig:system_model}, we consider a STAR-RIS-assisted UVAA communication system. Specifically, a set of UAVs denoted as $\mathcal{M}\triangleq \{1,\ldots, m,\ldots, N_M\}$ collaborates with a STAR-RIS fixed on a building to provide downlink communications to mobile ground users. These ground users can be divided into two categories: those on the same side of the STAR-RIS as the UAV swarm, denoted as $\mathcal{K}\triangleq \{1,\ldots, k,\ldots, N_K\}$, and those on the opposite side, denoted as $\mathcal{J}\triangleq \{1,\ldots, j,\ldots, N_J\}$. 

\par In the communication process, the STAR-RIS, which consists of a uniform planar array (UPA) with elements $\mathcal{S}\triangleq \{1,\ldots, s,\ldots, N_S\}$, receives and reconfigures signals from the UAVs, then simultaneously reflects these signals to same-side users $\mathcal{K}$ and transmits signals to opposite-side users $\mathcal{J}$. Moreover, it is considered that the UAVs and STAR-RIS can be controlled by a high-performance controller (\textit{e.g.}, field programmable gate arrays).

\par We consider that all the UAVs are equipped with omnidirectional single antennas and achieve synchronization in carrier frequency, time, and initial phase~\cite{Alemdar2021}. Additionally, data sharing among UAVs is completed by using the methods in~\cite{Feng2013}, while the channel state information (CSI) is obtained using the method in~\cite{Ahmad2022}. Without loss of generality, we model the system by using a three-dimensional (3D) Cartesian coordinate system. In this model, the position of the STAR-RIS, which remains constant over time, is denoted as $(x_\mathrm{S},y_\mathrm{S},z_\mathrm{S})$. Furthermore, the positions of the $m$th UAV, $k$th same-side user, and $j$th opposite-side user in the time slot $l \in \mathcal{L}=\{1,\ldots,l,\ldots, L\}$ are denoted as ${(x_m(l),y_m(l),z_m(l))}$, ${(x_k^R(l),y_k^R(l),0)}$, and ${(x_j^T(l),y_j^T(l),0)}$, respectively.

%
%
\subsection{STAR-RIS Model for ORB}
\label{ssec:Passive Beamforming}

\par Different from conventional RISs that can only reflect incident signals, the STAR-RIS breaks the conventional half-space coverage limitation by allowing electromagnetic waves to simultaneously pass through and reflect from its surface, thereby achieving \ang{360} full-space coverage~\cite{Yadav2023}. Therefore, this dual reflection-transmission capability enables more flexible ORB that can serve users on both sides of the surface, which is particularly advantageous in complex urban environments with UAV communications. In our model, we adopt the energy splitting (ES) protocol where each element of the STAR-RIS divides the incident signal energy into transmitted and reflected components with adjustable ratios.

\par For the $l$th time slot, we denote the reflection coefficient matrix as $\bm{\Theta}^R{(l)} \in \mathbb{C}^{N_s \times N_s}$ and transmission coefficient matrix as $\bm{\Theta}^T{(l)} \in \mathbb{C}^{N_s \times N_s}$:
\begin{equation}
    \bm{\Theta}^{R}(l)\triangleq \mathrm{diag}\left(\sqrt{a_{1}^{R}(l)}e^{j\theta_{1}^{R}(l)},\ldots,\sqrt{a_{N_{s}}^{R}(l)}e^{j\theta_{N_{s}}^{R}(l)}\right), 
\end{equation}
\begin{equation}
    \bm{\Theta}^{T}(l)\triangleq \mathrm{diag}\left(\sqrt{a_{1}^{T}(l)}e^{j\theta_{1}^{T}(l)},\ldots,\sqrt{a_{N_{s}}^{T}(l)}e^{j\theta_{N_{s}}^{T}(l)}\right), 
\end{equation}

\noindent where $a_{s}^{R}(l)$ and $a_{s}^{T}(l)$ represent the ES coefficients for reflection and transmission, respectively, with $a_{s}^{R}(l), a_{s}^{T}(l) \in [0,1]$. Moreover, the phase shifts $\theta_{s}^{R}(l)$ and $\theta_{s}^{T}(l)$ can be independently adjusted within $[0,2\pi)$. Following the energy conservation principle and assuming lossless elements, we have $a_{s}^{R}(l) + a_{s}^{T}(l) = 1$ for each element $s$.

\par Accordingly, by intelligently configuring these amplitude and phase parameters, the STAR-RIS can implement ORB. Specifically, the STAR-RIS is capable of simultaneously enhancing signal reception for intended users and mitigating interference to unintended receivers in both transmission and reflection modes.

%
%
\subsection{UVAA Model for CB}
\label{ssec:vaa_model}
\par In a UVAA, each UAV operates as an independent antenna element within the array, collectively performing CB through precise spatial positioning. Different from ORB with STAR-RIS, CB actively coordinates multiple transmitting entities to create constructive interference patterns. As a result, the electromagnetic waves transmitted by these UAVs can either superimpose or offset, thereby leading to high gain in certain directions. To evaluate the distribution of these signals mathematically, we use the array factor, which can be given by
\begin{equation}
\label{eq:af}
AF(\theta, \phi)=\sum_{m=1}^{N_M} I_m e^{j\left[k_c\left(x_m \sin \theta \cos \phi+y_m \sin \theta \sin \phi+z_m \cos \theta\right)\right]},
\end{equation}
\noindent where $\theta \in [ 0,\pi ]$ and $\phi \in[-\pi,\pi ]$ are the horizontal and vertical angles, respectively. Meanwhile, $I_m$ represents the excitation current weight of the $m$th UAV. Additionally, $k_c = 2\pi/\lambda$ is the phase constant, and $\lambda$ is the wavelength. By dynamically adjusting their 3D positions, the UAVs collaboratively shape the radiation pattern to maximize signal strength at the receiver while minimizing interference in other directions.

%
%
\subsection{Channel Model} 
\label{ssec:channel_model}

\par During wireless transmission, signal propagation is affected by various factors, including the path loss, multipath effects, and directivity of antennas. Therefore, as shown in Fig.~\ref{Fig:system_model}, the channel model of our system consists of three distinct link types. Specifically, the system incorporates a channel from UVAA to STAR-RIS, a direct channel from UVAA to the user, and a channel from STAR-RIS to the user. These links exhibit significantly different channel characteristics due to variations in the propagation environment. In the following, we establish corresponding parametric channel models for each type of channel and conduct a detailed analysis.

%
%
\subsubsection{Channel from UVAA to STAR-RIS} 
\par Given that the UVAA operates at a relatively high altitude and maintains a stable LoS connection with the STAR-RIS, the signal propagation is primarily affected by free-space path loss rather than severe multipath fading. Consequently, the channel from UVAA to STAR-RIS can be modeled as an LoS channel, with its channel gain expressed as
\begin{equation}
    \begin{aligned}
        \bm{h}&_{\mathrm{MS}}{(l)} =  AF(\phi_{\mathrm S}(l),\theta_{\mathrm S}(l)) \sqrt{\rho d_{\mathrm{MS}}^{-2}{(l)}} \bm{h}_{\mathrm{MS}}^{\mathrm{LoS}}{(l)},
        \label{Eq_channel_vaa_irs}
    \end{aligned}
\end{equation}
\noindent where $\phi_{\mathrm S}(l)$ and $\theta_{\mathrm S}(l)$ represent the horizontal and vertical angles of departure (AoD) from UVAA to STAR-RIS in time slot $l$. Moreover, $ \rho $ is the path loss coefficient, $d_{\mathrm{MS}}(l)$ denotes the distance between the UVAA center and the STAR-RIS at time slot $l$ and another key component $\bm{h}_{\mathrm{MS}}^{\mathrm{LoS}}{(l)}$ is given by 
\begin{equation}
    \begin{aligned}
        \bm{h}_{\mathrm{MS}}^{\mathrm{LoS}}{(l)} \!&=\!  {[1, e^{-j\frac{2 \pi d_r}{\lambda_c}}\chi_l^{\mathrm{MS}}\varphi_l^{\mathrm{MS}}, \ldots, e^{-j \frac{2 \pi (N_{S}^\mathrm{R}-1) d_{\mathrm{r}}}{\lambda_c}  }\chi_l^{\mathrm{MS}}\varphi_l^{\mathrm{MS}}]^{\mathrm{T}} } \\
         &\otimes[1, e^{-j \frac{2 \pi d_c}{\lambda_c}} \varrho _l^{\mathrm{MS}}\varphi_l^{\mathrm{MS}}, \ldots, e^{-j \frac{2 \pi(N_{S}^\mathrm{C}-1) d_{\mathrm{c}}}{{\lambda_c} } }\varrho _l^{\mathrm{MS}}\varphi_l^{\mathrm{MS}}]^{\mathrm{T}},
    \end{aligned}
    \label{eq:vaa_irs_los}
\end{equation}
\noindent where $\lambda_c$, $d_r$, $d_c$, $N_{S}^\mathrm{R}$, and $N_{S}^\mathrm{C}$ denote the carrier wavelength, row separation, column separation, and the number of elements in one row and one column of the STAR-RIS, respectively. In addition, $\varphi_l^{{\mathrm{MS}}}$, $\chi_l^{{\mathrm{MS}}}$, and $\varrho _l^{\mathrm{\mathrm{MS}}}$ represent the sine value of vertical angle of arrival (AoA), the cosine and sine values of the horizontal AoAs of the signal from the UVAA to STAR-RIS in time slot $l$, respectively. Note that the calculations of these three parameters follow~\cite{Wei2021}.

%
%
\subsubsection{Channel from UVAA to User} 
\par In the considered scenario, the Rayleigh channel model is suitable for modeling signal propagation from the UVAA to the user, as this propagation is affected by obstacles. Therefore, the channel gain from the UVAA to the same-side users $\mathcal{K}$ can be expressed as
\begin{equation}
    \bm{h}_{\mathrm{MK}}{(l)} = AF(\phi_{k}(l),\theta_{k}(l))\sqrt{\rho d_{\mathrm{MK}}^{-\alpha_d}{(l)}}\tilde h,
    \label{eq:uvaa_userk_channel}
\end{equation}

\noindent where $d_{\mathrm{MK}}{(l)}$, $\alpha_d$, $\phi_{k}(l)$ and $\theta_{k}(l)$ denote the distance between the UVAA and the $\mathcal{K}$, the corresponding path loss exponent, the horizontal and vertical AoDs from the UVAA to the $\mathcal{K}$, respectively. Moreover, $ \tilde h $ is modeled by a zero-mean and unit-variance circularly symmetric complex Gaussian random variable. Similarly, the channel gain from the UVAA to the opposite-side users $\mathcal{J}$ can be expressed as
\begin{equation}
    \bm{h}_{\mathrm{MJ}}{(l)} = AF(\phi_{j}(l),\theta_{j}(l))\sqrt{\rho d_{\mathrm{MJ}}^{-\alpha_d}{(l)}}\tilde h.
    \label{eq:uvaa_userj_channel}
\end{equation}

%
%
\subsubsection{Channel from STAR-RIS to User} 

\par As for the signal propagation from the STAR-RIS to the user, since the transmission path has a relatively clear LoS, the Rician channel model can fully take into account the contribution of possible reflection paths to the signal. Accordingly, for the same-side users $\mathcal{K}$, the channel gain can be given by 
\begin{equation}
    \begin{aligned}
        \bm{h}&_{\mathrm{SK}}{(l)} = \\& \sqrt{\rho d_{\mathrm{SK}}^{-\alpha_r}{(l)}} \left(\sqrt{\frac{\beta}{1+\beta}} \bm{h}_{\mathrm{SK}}^{\mathrm{LoS}}{(l)}+\sqrt{\frac{1}{1+\beta}} \bm{h}_{\mathrm{SK}}^{\mathrm{NLoS}}{(l)}\right),
    \end{aligned}
    \label{Eq_channel_star-irs_userk}
\end{equation}
\noindent where $d_{\mathrm{SK}}(l)$, $\alpha_r$, $\bm{h}_{\mathrm{SK}}^{\mathrm{LoS}}{(l)} $ and $\bm{h}_{\mathrm{SK}}^{\mathrm{NLoS}}{(l)}$ denote the distance between the STAR-RIS and the $\mathcal{K}$, the corresponding path loss exponent, the LoS and Non-LoS components, respectively. Specifically, $\bm{h}_{\mathrm{SK}}^{\mathrm{LoS}}{(l)}$ can be expressed as follows:
\begin{equation}
    \begin{aligned}
        \bm{h}_{\mathrm{SK}}^{\mathrm{LoS}}{(l)} \!&=\! {[1, e^{-j\frac{2 \pi s_r}{\lambda_c}}\chi_l^\mathrm{SK}\varphi_l^\mathrm{SK}, \ldots, e^{-j \frac{2 \pi (N_{S}^{R}-1) d_{\mathrm{r}}}{\lambda_c}}\chi_l^\mathrm{SK}\varphi_l^\mathrm{SK}]^{\mathrm{T}} } \\
        &\otimes[1, e^{-j \frac{2 \pi d_c}{\lambda_c}} \varrho _l^\mathrm{SK}\varphi_l^\mathrm{SK}, \ldots, e^{-j \frac{2 \pi (N_{S}^{C}-1) d_{\mathrm{c}}}{{\lambda_c}}}\varrho_l^\mathrm{SK}\varphi_l^\mathrm{SK}]^{\mathrm{T}},
    \end{aligned}
    \label{eq:star-irs_userk_los}
\end{equation}

\noindent where $\varphi_l^{\mathrm{\mathrm{SK}}}$, $\chi_l^{\mathrm{\mathrm{SK}}}$, $ \varrho _l^{\mathrm{\mathrm{SK}}}$ represent the sine value of vertical AoD, the cosine and sine values of the horizontal AoDs of the signal from STAR-RIS to the $\mathcal{K}$ in time slot $l$, respectively. 
Additionally, the non-LoS component $ \bm{h}_{\mathrm{SK}}^{\mathrm{NLoS}}{(l)}$ of Eq.~\eqref{Eq_channel_star-irs_userk} is drawn independently from the circularly symmetric complex Gaussian distribution with zero mean and unit variance.
\par Likewise, the channel gain from the STAR-RIS to the opposite-side users $\mathcal{J}$ is given by
\begin{equation}
    \begin{aligned}
        \bm{h}&_{\mathrm{SJ}}{(l)} = \\& \sqrt{\rho d_{\mathrm{SJ}}^{-\alpha_r}{(l)}} \left(\sqrt{\frac{\beta}{1+\beta}} \bm{h}_{\mathrm{SJ}}^{\mathrm{LoS}}{(l)}+\sqrt{\frac{1}{1+\beta}} \bm{h}_{\mathrm{SJ}}^{\mathrm{NLoS}}{(l)}\right).
    \end{aligned}
    \label{Eq_channel_star-irs_userj}
\end{equation}

%
%

\subsection{Secure Transmission Models}
\label{ssec:transmission_model}
\par Considering the differences in spatial channels and the combined gains from direct, reflective and transmissive paths, we construct a differentiated signal reception model for users $\mathcal{K}$ and $\mathcal{J}$.

\par Specifically, the same-side users $\mathcal{K}$ receive the direct wave from UVAA and the reflected wave from STAR-RIS, while the opposite-side users $\mathcal{J}$ receive the direct wave from UVAA and the transmitted wave from STAR-RIS. Therefore, we can express the total gain for the $\mathcal{K}$ and $\mathcal{J}$ in time slot $l$ as follows:
\begin{equation}
    \begin{aligned}
        G_K(l)  = \frac{4\pi \left| (\bm{h}_{\mathrm{MS}}{(l)})^T\bm{\Theta}^R{(l)}\bm{h}_{\mathrm{SK}}{(l)}  +h_{\mathrm{MK}}{(l)} \right|^2}
         {{\int_0^{2 \pi} \int_0^\pi|AF(\theta, \phi)|^2 w(\theta, \phi)^2 \sin \theta \mathrm{d} \theta \mathrm{d} \phi}}\eta, 
    \end{aligned}
    \label{eq:gain1}
\end{equation}
\begin{equation}
    \begin{aligned}
        G_J(l)  = \frac{4\pi \left| (\bm{h}_{\mathrm{MS}}{(l)})^T\bm{\Theta}^T{(l)}\bm{h}_{\mathrm{SJ}}{(l)}  +h_{\mathrm{MJ}}{(l)} \right|^2}
         {{\int_0^{2 \pi} \int_0^\pi|AF(\theta, \phi)|^2 w(\theta, \phi)^2 \sin \theta \mathrm{d} \theta \mathrm{d} \phi}}\eta, 
    \end{aligned}
    \label{eq:gain2}
\end{equation}
\noindent where $w(\theta, \phi)$ represents the amplitude of the far-field beam pattern for each element, and $\eta \in [0,1]$ denotes the array efficiency. 
\par Following this, the total transmission rate of the system, can be defined as follows:
\begin{equation}
    R(l) = B\log_2(1+\frac{P_t G_K(l)}{\sigma^2}) + B\log_2(1+\frac{P_t G_J(l)}{\sigma^2}), 
    \label{eq:transmission_rate_total}
\end{equation}

\noindent where $B$, $P_t$, and $\sigma^2$ denote the transmission bandwidth, total transmit power of the UVAA, and noise power, respectively. 

%
%
\subsection{Energy Consumption Model of UAV} 
\label{ssec:energy_model}
\par According to~\cite{Zeng2019}, the total energy consumption of a rotary-wing UAV consists of two main components. The first component is related to communication functions, which includes the energy needed for electromagnetic wave transmission, signal processing, and circuit operation. The second component is the propulsion energy consumption for maintaining the flight state, which ensures hovering and maneuvering capabilities.

\par Note that the energy consumed for the communication of UAV is often negligible, as it is significantly lower than the energy required for propulsion, typically a few watts compared to hundreds of watts~\cite{Zeng2019}. Based on this observation, the power consumption for propulsion in a rotary-wing UAV flying at a speed $v$ in a two-dimensional (2D) plane can be modeled as follows:
\begin{equation}
    \begin{aligned}
        P(v)=& P_B\left(1+\frac{3v^2}{v_{t i p}^2}\right)+P_I\left(\sqrt{1+\frac{v^4}{4 v_0^4}}-\frac{v^2}{2 v_0^2}\right)^{1 / 2}+ \\
        & \frac{1}{2} d_0 \rho s A v^3,
    \end{aligned}
    \label{eq:uav_power}
\end{equation}
\noindent where $P_B$ and $P_I$ are the two constants that represent the blade profile and induced powers in the hovering status, respectively. Moreover, $v_{t i p}$ is the tip speed of the rotor blade, $v_0$ is the mean rotor-induced velocity in hovering, and $d_0$ and $s$ are the fuselage drag ratio and rotor solidity, respectively. Furthermore, $\rho$ and $A$ are known as the air density and rotor disc area, respectively. 
\par As indicated in~\cite{Gong2024}, since UAV horizontal flight acceleration and deceleration constitute only a minor portion of the UAV maneuver duration, we neglect the additional energy consumption increase or decrease caused by UAV horizontal flight acceleration and deceleration. Therefore, considering arbitrary 3D trajectories involving UAV ascent and descent, the heuristic approximate expression for UAV 3D flight energy consumption is given by
\begin{equation}
    \begin{aligned}
        E(l) \approx & P(v(l))\triangle l+\frac{1}{2} m_{M}\left(\bar v(l)^2-\bar v(l-1)^2\right)+\\
        & m_{M} g(z(l)-z(l-1)),
    \end{aligned}
    \label{eq:uav_energy}
\end{equation}
\noindent where $v(l)$, $m_M$, and $g$ are the instantaneous speed of the UAV at time slot $l$, the aircraft mass of UAV, and the gravitational acceleration, respectively.

%
\section{Problem Formulation and Analysis} \label{sec:formulation}

\par In this work, we aim to optimize the STAR-RIS-assisted UVAA communication system through jointly maximizing transmission rate of the overall system while minimizing energy consumption of the UAV swarm. However, these objectives present an inherent trade-off. Specifically, enhancing the transmission rate of the overall system typically requires aggressive signal processing, higher power allocation, and frequent STAR-RIS reconfigurations, all of which increase UAV energy consumption. Conversely, minimizing UAV energy consumption by reducing transmission power, simplifying signal modulation, or decreasing STAR-RIS adjustment frequency inevitably constrains achievable transmission rates. This trade-off necessitates a balanced strategy that prevents either objective from being overly compromised. 

\par To elaborate, based on the system model, the total transmission rate is jointly determined by the trajectories and excitation current weights of UVAA, the transmission and reflection matrices of STAR-RIS, and the time-varying channel power gains. Furthermore, the energy efficiency of UAVs is closely related to the UAV flight trajectories during the mission period. As such, we let $\bm{I} = \{I_m(l) \mid m \in \mathcal{M}\} $, $ \bm{v} = \{v_m(l) \mid m \in \mathcal{M}\}$, $ \bm{\phi} = \{\phi_m(l) \mid m \in \mathcal{M}\}$ and $\bm{\omega} =\{\omega_m(l) \mid m \in \mathcal{M}\}$ denote the set of excitation current weights, horizontal speed set, horizontal flight direction and vertical speed of all UAVs, respectively. Based on the aforementioned parameter definitions and considering both the total system transmission rate and UAV energy consumption, the optimization objective of the system can be formulated as
\begin{equation}
    \eta(l)=\lambda_1 R(l)-\lambda_2 E(l),
\end{equation}
\noindent where $\lambda_1 $ and $ \lambda_2 $ are the weighting coefficients that quantify the emphasis on transmission rate and energy consumption, respectively. 

\par Following this, the considered optimization problem JREOP can be formulated as follows:
\begin{subequations}
    \label{eq:problem}
    \begin{align}
    \max _{\mathbf{\Theta}^R(l), \mathbf{\Theta}^T(l), \bm{I}, \bm{v}, \bm{\phi}, \bm{\omega}} \ &\sum_{l=1}^{N_L} \eta(l) ,\\
    \mathrm{s.t.}\quad\quad\quad
    &0 \le I_m{(l)} \le 1,\forall m \in \mathcal{M}, \label{conji}\\ 
    &L_{\min } \leqslant x_m(l) \leqslant L_{\max }, \forall m \in \mathcal{M}, \label{conx} \\
    &L_{\min } \leqslant y_m(l) \leqslant L_{\max }, \forall m \in \mathcal{M}, \label{cony} \\
    &H_{\min } \leqslant z_m(l) \leqslant H_{\max }, \forall m \in \mathcal{M}, \label{conz}\\
    &D_{(m 1, m 2)} \geq D_{\min }, \forall m_1, m_2 \in \mathcal{M}, \label{collision} \\
    &B \log _2\left(1+\frac{P_t G_{\mathrm{K}}(l)}{\sigma^2}\right) \geq R_{K, \min }, \label{conk}\\
    &B \log _2\left(1+\frac{P_t G_{\mathrm{J}}(l)}{\sigma^2}\right) \geq R_{J, \min }, \label{conj}\\
    &\theta_s^R(l), \theta_s^T(l) \in[0,2 \pi), \forall s \in \mathcal{S}, \label{conS1}\\
    &a_s^R(l), a_s^T(l) \in[0,1], \forall s \in \mathcal{S}, \label{conS2}\\
    &a_s^R(l)+a_s^T(l)=1, \forall s \in \mathcal{S}, \label{conS3}
    \end{align}
\end{subequations}
\noindent where $ L_{min} $ and $ L_{max} $ are the minimum and maximum ranges of the area that the UAVs can move in the horizontal plane, and $ H_{min} $ and $ H_{max} $ are the minimum and maximum altitudes that the UAVs can fly in the vertical directions, respectively. The constraint in \eqref{collision} indicates that the minimum distance between two neighboring UAVs must be greater than $D_{min}$ to avoid collisions. Moreover, \eqref{conk} and \eqref{conj} ensure that the corresponding users meet the minimum data rate requirements. Finally, \eqref{conS1}, \eqref{conS2}, and \eqref{conS3} ensure that the amplitude and phase shift coefficients of the STAR-RIS will be adjusted within reasonable ranges. 

\par In addition to its mathematical non-convexity and NP-hardness, JREOP also exhibits several structural characteristics. In the following, we further analyze JREOP from three perspectives, \textit{i.e.}, its dynamic nature, structural heterogeneity, and long-term objective orientation. \textit{First}, the problem possesses high dynamic characteristics because it encompasses evolving user trajectories, time-varying channel states, and stochastic initial positions of UAVs. Therefore, this problem requires solutions that can adapt to rapidly changing deployment conditions. \textit{Second}, JREOP exhibits significant heterogeneity in the decision space, which simultaneously involves UAV swarm optimization and STAR-RIS configuration. These two subsystems have distinct action spaces but are strongly interdependent. Specifically, the motion decisions of UAVs affect the optimal STAR-RIS configuration, while the STAR-RIS transmission and reflection matrices in turn influence communication quality and UAV positioning. \textit{Third}, JREOP is oriented toward a long-term cumulative objective function that evaluates performance over an extended time horizon rather than instantaneous optimization. This orientation necessitates solutions that balance immediate rewards with future consequences of current decisions. These properties collectively highlight the necessity of developing an adaptive and robust optimization framework. 

\par Based on these requirements, MADRL provides an ideal framework. To begin with, MADRL addresses dynamicity through adaptive decision-making via continuous agent-environment interaction. In addition, MADRL handles heterogeneity effectively by modeling diverse system components as independent yet collaborative agents with specialized learning capabilities. Moreover, the focus of MADRL on maximizing cumulative rewards naturally aligns with our need for long-term performance optimization and helps to balance immediate gains against future benefits. Therefore, we propose an MADRL-based approach to solve the JREOP in the following sections.

\section{Our Heterogeneous Multi-Agent Collaborative Dynamic Optimization Framework} \label{sec:algorithm}

\par In this section, we first reformulate the JREOP as a heterogeneous MDP. Subsequently, we introduce the ATSO for controlling the STAR-RIS agent. Building upon this, we propose an MADRL-based algorithm for coordinating the UAV swarm. Finally, we summarize our proposed HMCD framework.

%
\subsection{Heterogeneity MDP Formulation}

\par In the system, we consider two heterogeneous agent categories, specifically UAVs and STAR-RIS, each with distinct observation spaces and action capabilities. To systematically address this heterogeneity, we reformulate the optimization problem presented in Eq.~\eqref{eq:problem} within the framework of a heterogeneous MDP. Formally, this MDP is characterized by the quintuple $(\mathcal{S}, \mathcal{A}, \mathcal{P}, \mathcal{R}, \gamma)$, where $\mathcal{S}$ represents the state space, $\mathcal{A}$ denotes the action space, $\mathcal{P}$ signifies the state transition probability distribution, $\mathcal{R}$ corresponds to the reward function, and $\gamma \in [0,1)$ serves as the discount factor governing the tradeoff between immediate and future rewards. Subsequently, we detail the design of the state space, action space, and reward within our model. 

%
\subsubsection{State Space} 

\par In practical implementations, CB necessitates robust data sharing and synchronization protocols among UAVs, thereby requiring each UAV to precisely monitor the positions of neighboring UAVs through established methods presented in~\cite{Alemdar2021}. For modeling user mobility patterns, this study adopts the Gauss-Markov random movement model (GMRMM) as delineated in~\cite{yang2022onlineGMRMM}, with all associated parameters configured in accordance with the specifications therein. Furthermore, considering the requisite environmental information exchange between the UAV network and the STAR-RIS, the system state of the wireless communication architecture, jointly optimized through UVAA and STAR-RIS, at discrete time slot $l$ is formally characterized as follows:
\begin{equation}
    \begin{aligned}
        \bm s_{l} = &\{x_m(l), y_m(l), z_m(l) \mid m \in \mathcal{M}\} \\ & \cup \{x_k(l),y_k(l)\} \cup \{x_j(l),y_j(l)\}.
    \end{aligned}
    \label{eq:state}
\end{equation}

%
\subsubsection{Action Space}

\par Both UVAA and STAR-RIS need to dynamically adjust their respective parameters, so as to respond to dynamic channel conditions and satisfy communication requirements. Specifically, after obtaining the environmental state, each UAV agent will take actions to adjust its excitation current weight, horizontal speed, horizontal flight direction, and vertical speed. These actions can be denoted by $\bm {a}_l^m = \{{I}_m(l), v_m(l), \phi_m(l), \omega_m(l)\}$ and are determined based on the state of UVAA, the motion characteristics of neighboring UAVs, and the configuration of the STAR-RIS. As such, the action space for all the UAVs is given by
\begin{equation}
    \begin{aligned}
    \bm{a}_l= \left\{ \bm{a}_l^m \mid m \in \mathcal{M} \right\}.
    \label{eq:UAV_action}
    \end{aligned}
\end{equation}

\par Meanwhile, the STAR-RIS agent performs amplitude and phase adjustments for both transmission and reflection to enable beamforming and channel optimization, and the action is given by
\begin{equation}
    \bm{a}_l^\mathrm{S} =\bm{\Theta}^R{(l)} \cup \bm{\Theta}^T{(l)}.
\end{equation}

%
\subsubsection{Reward Design} 

\par All agents share the common objective defined in Eq.~\eqref{eq:problem}, which aims to maximize system transmission rate while minimizing energy consumption. To this end, the proposed reward structure integrates the timeslot-specific objective function $\eta_m(l)$ with directional guidance terms to promote efficient exploration of the action space. Specifically, these guidance terms consist of \textit{STAR-RIS proximity guidance}, which encourages UAVs to move toward the STAR-RIS, and \textit{boundary avoidance}, which penalizes trajectories that approach the edge of the operational region. Additionally, strategic penalty terms enforce constraints from expressions \eqref{conx}-\eqref{collision}. 

\par Accordingly, the reward function $r_l^m$ for agent $m$ at time slot $l$ is formulated as follows:
\begin{equation}
r_{l}^{m} = 
\begin{cases}
\eta_{m}(l) 
+ \zeta_1\frac{\bm{q}_l^{m,\mathrm{R}}\cdot {\bm {z}_l^m}} {\Vert{\bm{q}_l^{m,\mathrm{R}}}\Vert \Vert {\bm {z}_l^m} \Vert}
- \zeta_2 \Vert{\bm{q}_l^{m,\xi}}\Vert, & \text{if feasible} \\
\\
-\left(\epsilon_{1} O_{m}^{1} + \epsilon_{2} \sum_{m^{\prime} \in \mathcal{M}} C_{m m^{\prime}}^{2}\right), & \text{otherwise}
\end{cases}
\label{eq:reward}
\end{equation}

\noindent where the term \textit{feasible} indicates that all constraints defined in Eqs.~\eqref{conx}–\eqref{collision} are satisfied. In addition, $\bm{q}_l^{m,\mathrm{R}}$ denotes the vector pointing from the UAV to the STAR-RIS, $\bm{z}_l^m$ represents the displacement vector of the UAV, and $\bm{q}_l^{m,\xi}$ corresponds to the vector from the UAV to a predefined reference point $\bm{\xi}$ located within the feasible flight region. Moreover, the parameters $\zeta_1$ and $\zeta_2$ adjust the relative emphasis between direction following and boundary avoidance.

\par Furthermore, $\epsilon_1$ and $\epsilon_2$ are dynamically increasing penalty weights to penalize UAVs for boundary violations and collisions, respectively. Specifically, both weights follow the formula $\epsilon_i = \epsilon + (1 - \epsilon) \times \min(t/{t_{\text{max}}}, 1.0)$, where $t$ represents the current timestep and $t_{max}$ is the maximum mission duration. Moreover, $O_m^1$ and $C_{mm'}^2$ are two binary variables indicating whether the UAV $m$ is out of the boundary and whether it collides with the UAV $m^\prime$, respectively.

\par Evident from the preceding analysis, the UAVs and STAR-RIS agents constitute a heterogeneous multi-agent system, characterized by distinct action spaces while functioning collaboratively to optimize the cumulative reward objective. This heterogeneity in agent functionality necessitates the development of differentiated control policies. Therefore, in the subsequent sections, we present specialized policy frameworks for STAR-RIS and UAVs.

%
\subsection{ATSO-Driven Control of the STAR-RIS}
\label{subsection:irs_control_policy}

\par The STAR-RIS system faces significant optimization challenges in dynamic environments, including high-dimensional parameter spaces and stringent real-time optimization requirements. To address these challenges, we propose the ATSO algorithm, which transforms the multi-slot STAR-RIS control problem into single-slot optimization while employing simulated annealing (SA) to improve exploration of the solution space~\cite{Huang2024}~\cite{Xia2024}.

\par Specifically, the SA utilizes a temperature parameter $T$ to balance exploration and exploitation, which can accept suboptimal solutions with probability $P(\text{accept}) = \exp\left({\Delta E}/{T}\right)$ during early phases, where $\Delta E$ represents the change in objective function value. As optimization progresses, temperature gradually decreases according to a cooling schedule $T_{k+1} = \alpha \cdot T_k$, where $\alpha < 1$ is the cooling rate.

\par Consequently, our proposed ATSO begins with temperature initialization and proceeds through several key steps. \textit{First}, generating adaptive candidate coefficient sets based on current temperature. \textit{Second}, evaluating performance metrics for all candidates. \textit{Third}, selecting candidates probabilistically using temperature-scaled distributions when $T > T_{\text{min}}$ or deterministically when $T \leq T_{\text{min}}$. \textit{Finally}, updating STAR-RIS matrices accordingly.

\par Notably, the ATSO generates optimized STAR-RIS configurations that directly influence channel conditions and signal propagation paths, which in turn affect the reward signals and state observations for the MADRL component. Conversely, the UAV positions and trajectories determined by MADRL create dynamic channel states that continually reshape the optimization landscape for ATSO. This bidirectional relationship establishes an adaptive control loop where the reflection and transmission coefficients of ATSO evolve in response to UAV movements, while UAV coordination strategies adapt to the changing electromagnetic environment created by STAR-RIS configurations.

\begin{algorithm}[tb]
\caption{ATSO}
\label{alg:star_ris_sa}
\KwIn{Initial temperature $T_{\text{init}}$, cooling rate $\alpha$, minimum temperature $T_{\text{min}}$}
Initialize $T \leftarrow T_{\text{init}}$\;
\For{each element $i$ in STAR-RIS}{
    Define adaptive candidate sets based on temperature $T$\;
    Generate all possible combinations of amplitude and phase candidates\;
    Compute $R_{\text{candidates}} = \sqrt{a} \cdot e^{j\theta_R}$ and $T_{\text{candidates}} = \sqrt{1-a} \cdot e^{j\theta_T}$\;
    Compute current signal contributions $S_k$ and $S_j$\;
    Calculate $\text{current\_metric} = |S_k + S_j|^2$\;
  
    Calculate $\text{metrics}$ for all candidates\;
    Find $\text{best\_idx} = \text{argmax}(\text{metrics})$\;
  
    \If{$T > T_{\text{min}}$}{
        Calculate $\text{prob}$\;
        Select candidate using multinomial sampling from prob\;
    }
    \Else{
        Select candidate using best\_idx\;
    }
  
    Update $R[i]$ and $T[i]$ with selected candidate values\;
    Update temperature $T \leftarrow \max(\alpha \cdot T, T_{\text{min}})$\;
}
\KwOut{Optimized reflection matrix $R$ and transmission matrix $T$}
\end{algorithm}

%
\subsection{MADRL-Driven Control of the UVAA}

\par In the considered framework, we model each UAV as an autonomous agent and employ MADRL to enable cooperative behavior. Among various MADRL approaches~\cite{haarnoja2018SAC}~\cite{pmlr-v80-haarnoja18bMADDPG}~\cite{Cheng2025}, we adopt MASAC as the baseline algorithm, mainly because its maximum-entropy formulation improves exploration efficiency and reduces the need for excessive sampling. Based on this choice, the following sections provide a structured description of our methodology. Specifically, we first outline the theoretical basis of MASAC, after which we introduce two improvements, namely a self-attention evaluation mechanism for inter-agent coordination and an adaptive velocity transition mechanism for stable learning.

\subsubsection{Preliminaries of MASAC}
\label{subsubsection:MASAC}

\par MASAC extends the standard reinforcement learning algorithm by incorporating maximum entropy principles and state value estimations to enhance exploration capabilities. The fundamental objective of MASAC is to train a policy $\pi_{\theta_m}(\bm a_l^m\mid \bm s_l)$ that maximizes not only the expected cumulative rewards but also the entropy of the policy at each encountered state. This objective can be formally expressed as
\begin{equation}
\label{eq:best_pi}
    \pi_{\theta_m}^* = \arg \max _{\pi_{\theta_m}} \sum_{l=1}^L \mathbb{E}_\pi[r_l^m+\alpha \mathcal{H}(\pi_{\theta_m}(\cdot \mid \bm{s}_l))],
\end{equation}

\noindent where $\alpha$ represents the temperature parameter that governs the relative importance of the entropy term $\mathcal{H}(\pi_{\theta_m}(\cdot \mid \bm{s}_l))$ within the optimization framework. Following this, the soft target value $y$ can be obtained by
\begin{equation}
\label{eq:target}
    y= r_l^m +\gamma \mathbb{E}[Q_{\bar{\psi}_m} \left(\bm{s}_{l+1}, {\bm{a}_{l+1}}\right)-\alpha\log\pi_{\bar{\theta}_m}(\bm{a}_{l+1}^m\mid \bm{s}_{l+1})],
\end{equation}

\noindent where $Q_{\bar{\psi}_m}$ denotes the target value function and $\bm{a}_{l+1} = \{\bm a_{l+1}^m \sim \bar \theta _m(\cdot \mid \bm s_{l+1})\mid m \in \mathcal{M}\}$ represents the collection of target actions~\cite{haarnoja2018SAC}. Consequently, the parameters of the soft Q-function undergo updates through temporal-difference learning by minimizing the following objective function:
\begin{equation}J_Q\left(\psi_m\right)=\sum_{m=1}^{N_M}\mathbb{E}_{(\bm{s}_{l},\bm{a}_l,r_l,\bm{s}_{l+1}) \sim \mathcal{B}} \left[\dfrac{1}{2}(Q_{\psi_m}\left({\bm{s}_l}, {\bm{a}_l}\right) - y)^2\right].
    \label{cr_update}
\end{equation}

\par Furthermore, the policy parameters $\theta_m$ are optimized via stochastic gradient descent to iteratively minimize the magnitude of the following loss function, \textit{i.e.},
\begin{equation}
    \begin{aligned}
        J_\pi(\theta_m) =\mathbb{E}_{(\bm{s}_l,\bm{a}_l)\sim \mathcal{B},\bm{a}_l^m\sim \pi_{\theta_m}(\cdot\mid \bm{s}_l)} &
        [ \alpha  \log \pi_{\theta_m}(\bm{a}_l^m \mid \bm{s}_l) \\ & -Q_{{\psi}_m}(\bm{s}_l,\hat{\bm a})],
    \end{aligned}
\label{ac_update}
\end{equation} 	
\noindent where $\bm{a}_l^m$ represents the action sampled from the current policy for gradient computation. Within the composite action vector $\hat{\bm a}$, the component corresponding to agent $m$ is set to $\bm{a}_l^m$, while the actions for other agents are sampled from the experience replay buffer $\mathcal{B}$.

\par To enhance training stability, the target critic and actor networks are updated through a soft update mechanism:
\begin{equation}
    \begin{aligned}
         \bar{\psi}_m=\tau \psi_m+(1-\tau) \bar{\psi}_m,\quad
         \bar{\theta}_m=\tau \theta_m+(1-\tau) \bar{\theta}_m,
    \end{aligned}
\label{tar_update}
\end{equation}

\noindent where $\tau$ denotes the soft update coefficient controlling the rate of target network updates.

\par Despite its advantages, the application of MASAC to our specific JREOP reveals several critical limitations. \textit{First}, the complex interdependencies among variables introduce significant challenges for critic networks in accurately evaluating states, potentially leading to suboptimal policy convergence. \textit{Second}, negative early training experiences (like collisions and boundary violations) hinder the ability of agents to develop optimal velocity profiles. These identified limitations motivate our subsequent algorithmic enhancements.

\begin{figure*}[t]
    \centering
    \includegraphics[width=1.0\linewidth]{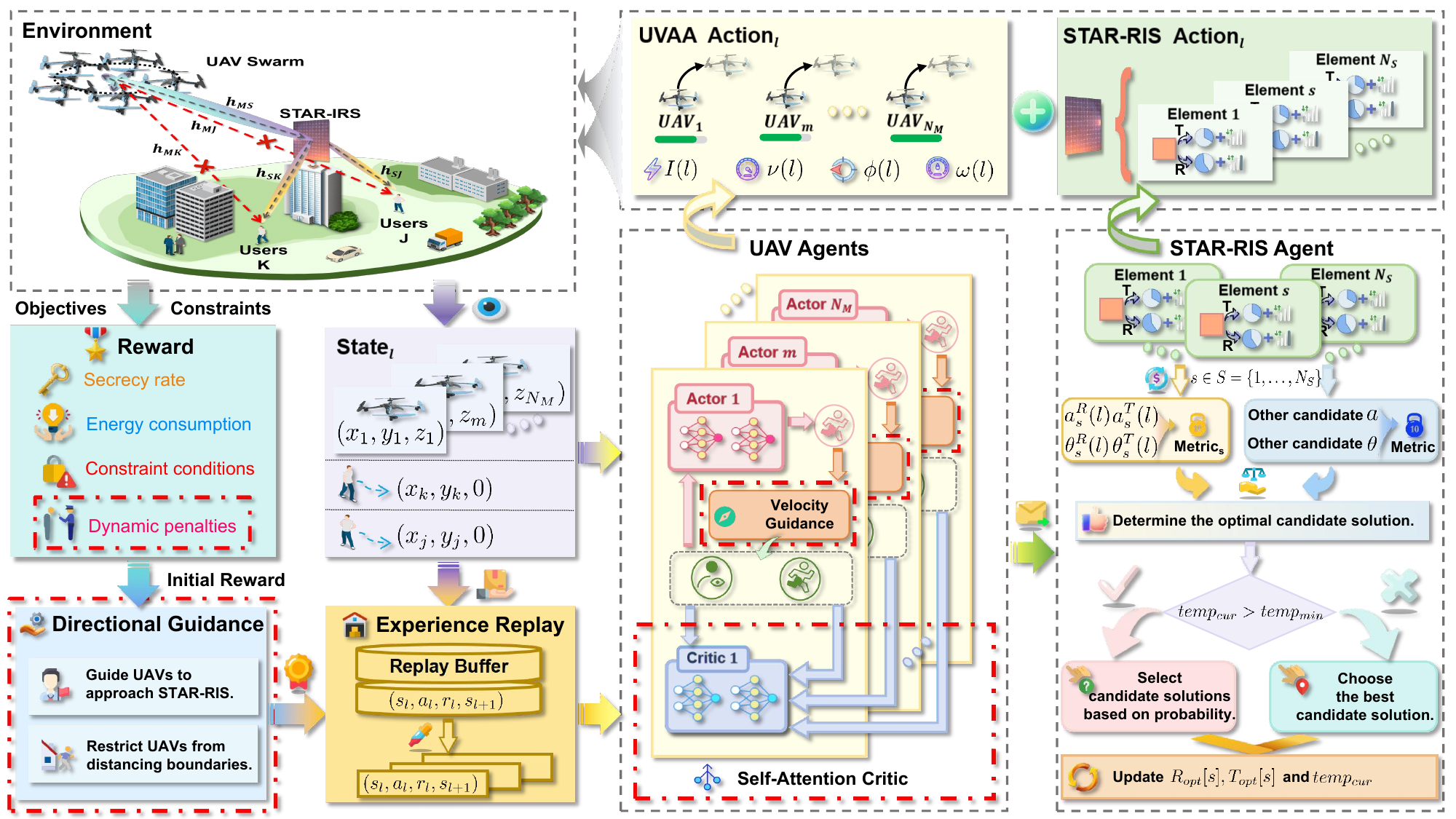}
    \caption{The proposed HMCD structure has two core components: an ATSO strategy for STAR-RIS control and an MADRL method for UAV coordination. The MADRL method features two key improvements: a self-attention evaluation mechanism and an adaptive velocity transition mechanism.}
    \label{Fig:HMCD}
\end{figure*}

%
\subsubsection{Collaboration-Aware Self-Attention Critic Mechanism}

\par In our formulated MDP framework, UAV agents operate within a cooperative paradigm characterized by non-conflicting objectives. Within such collaborative environments, we observe that conventional multilayer perceptron (MLP) architectures exhibit inherent limitations in capturing the complex inter-agent interaction dynamics~\cite{vaswani2017attention}. To address this critical limitation, we propose the integration of a self-attention mechanism within the critic network architecture, which enhances the capacity of network for collaborative reward evaluation by explicitly modeling agent interdependencies. 
\par Specifically, our attention-augmented Q-function $Q_{\psi_m}(\bm{s}, \bm{a})$ synthesizes information across agents through weighted attention to the actions of other UAVs, which facilitates more accurate value estimation in scenarios requiring coordinated behavior. The mathematical formulation of this attention-based value function is given by
\begin{equation}
    Q_{\psi_m}(\bm{s}, \bm{a}) = f_m(g_m(\bm{s}, \bm{a}_l^m), \bm{x}_m),
    \label{eq:attention_Q}
\end{equation}
\noindent where $f_m$ is a two-layer MLP and $g_m$ is a one-layer MLP, while $\bm{x}_m$ is the attention interaction information of the agent $m$ with other agents. To obtain $\bm{x}_m$, we use a set of parameters $\bm{W}_q$, $\bm{W}_k$ and $\bm{W}_v$ to accomplish the relevant transformations. According to~\cite{vaswani2017attention}, the query $q_m$, key $k_m$ and value $v_m$ can be expressed as
\begin{equation}
    q_m = g_m(\bm{s},\bm{a}_l^m)\bm{W}_q, \ k_m= \bm{a}_l^m  \bm{W}_k, \ v_m=  \bm{a}_l^m\bm{W}_v.
    \label{eq:attention_query}
\end{equation}

\par Then, the attention score of the agent $m$ to the agent $n$ is given by $s_{m,n} = \mathrm{softmax}(q_m k_n^T/\sqrt{d_{k}})$, where $ d_{k} $ is the dimension of the $k_m$. As such, $ \bm{x}_m $ can be obtained by $\bm x_m = \sum_{m \ne n } s_{m,n} v_n$.

\par Note that the critic network is not involved during execution, which means that our method still maintains the advantages of decentralization.

%
\subsubsection{Adaptive Velocity Transition Mechanism}

\par Training multi-agent UAV systems presents notable challenges due to frequent boundary constraint violations and inter-agent collisions during initial learning phases. Such operational violations prevent the flight trajectories from converging to energy-efficient patterns. Leveraging well-established energy optimization principles for aerial vehicles~\cite{Zeng2019}, we recognize that optimal energy consumption occurs at a particular velocity point, which we denote as $v_\mathrm{me}$. This optimal cruising speed can be determined analytically based on the aerodynamic parameters that characterize each vehicle type.

\par Building on this insight, we propose an adaptive velocity transition mechanism that balances energy efficiency with exploration. Our method employs a training-dependent interpolation mechanism that initially guides agents toward energy-optimal velocities~\cite{Wei2024}, while over time the policy gradually assumes greater autonomy. This adaptive guidance promotes stability during initial learning and enables the development of more nuanced strategies in later stages.

\par Accordingly, our adaptive velocity transition mechanism progressively transforms the policy output into an operational velocity through a training-dependent interpolation. Specifically, at training step $n$, the adjusted velocity $\tilde{v}_m$ is determined by
\begin{equation}
\label{eq:velocity_adjustment}
    \tilde{v}_m = \max\left(v_{min}, \min\left(v_{max}, \zeta v_m + (1-\zeta)v_b\right)\right),
\end{equation}

\noindent where the interpolated component combines the raw policy output $v_m$ with a guidance velocity $v_b \sim \mathcal{N}(v_{\mathrm{me}}, \sigma_b^2)$. The interpolation factor $\zeta = {n}/N_{E}$ gradually increases with training progress, transitioning from energy-efficient guidance toward full policy control while ensuring operation within the velocity bounds $[v_{min},v_{max}]$.

\begin{algorithm}[t]
    \KwIn{train episodes $N_{E}$, update times $N_{U}$.}
    \LinesNumbered %
    \caption{HMCD}
    \label{alg:Algorithm_HMCD} 
    \textbf{Initialization:} Initialize the actor $\theta_m$, target actor $\bar{\theta}_m =\theta_m$, critic $\psi_m$, target critic $\bar{\psi}_m = \psi_m$ and replay buffer $\mathcal{B}$ for each agent;\\
    \For{$ n=1 $ to $N_E$}
    {
        Initialize and observes the state $\bm{s}_l$;\\
        \For{$ l=1 $ to $N_L$}
        {
            Each UAV sample obtained $\bm{a}_l^m$ by sampling distribution $ \pi_{\theta_m}(\cdot\mid \bm{s}_l) $;\\
            Add gravity noise to $\bm{a}_l^m$ using Eq.~\eqref{eq:velocity_adjustment} and obtain the noised action $\hat{\bm a}_l^m$;\\
            Obtain the optimized reflection matrix $\bm{R}$ and transmission matrix $\bm{T}$ of the STAR-RIS by Algorithm~\ref{alg:star_ris_sa};\\
            UAVs and the STAR-RIS perform actions and receive the reward $r_l^m$;\\
            Obtain the next state $\bm{s}_{l+1}$;\\
            Union and store the experience tuple $(\bm{s}_l, \hat{\bm{a}}_l,{\bm r}_l,\bm{s}_{l+1})$ in $\mathcal{B}$;\\
            Update state $\bm s_{l} \leftarrow \bm s_{l+1}$;\\
            
            \For{$i=1$ to $N_{U}$}
            {
                Randomly sample a mini-batch of experiences $ (\bm{s},\bm{a},\bm{r},\bm{s}^\prime) $ from $ \mathcal{B} $;\\        
                \For{$m=1$ to $N_{M}$}{
                    Calculate $Q_{\psi_m}(\bm s,\bm a)$ and update $\psi_m$ using Eqs.~\eqref{cr_update} and~\eqref{eq:attention_Q};\\
                    Update $\theta_m$ using Eq.~\eqref{ac_update};\\
                    Update $\bar \psi_m$ and $\bar \theta_m$ using Eq.~\eqref{tar_update};\\
                }
            }
        }
    }		
    \KwOut{Policies $\{\pi_{\theta_1},\ldots,\pi_{\theta_m},\ldots,\pi_{\theta_{N_M}}\}$;\\}
\end{algorithm}

\subsection{Main Structure and Main Steps}

\par Based on the methods introduced above, we propose a HMCD framework to address the optimization problem formulated in Eq.~\eqref{eq:problem}. The overall structure of HMCD is depicted in Fig.~\ref{Fig:HMCD}, while the detailed procedure is outlined in Algorithm~\ref{alg:Algorithm_HMCD}.

\par As shown in Fig.~\ref{Fig:HMCD}, HMCD integrates the control of both the STAR-RIS and the UAV swarm within a unified decision-making architecture to achieve intelligent and efficient communication scheduling. In this framework, the STAR-RIS agent operates based on the control policy ATSO, while the UAV agents are governed by an enhanced MASAC approach. To support this design, each UAV agent incorporates a quadruple neural network architecture characterized by parameter sets $\theta_m$, $\psi_m$, $\bar \theta_m$, and $\bar \psi_m$, corresponding to policy, critic, target policy, and target critic networks, respectively.

\par During the training phase, both the STAR-RIS and UAV agents interact with the environment and store experience tuples in a shared replay buffer $\mathcal{B}$. Specifically, each UAV agent first generates a preliminary action $a_l^m$ via its policy network, which is subsequently transformed through the adaptive velocity
transition mechanism defined in Eq.~\eqref{eq:velocity_adjustment} to produce the exploration-enhanced action $\hat{\bm a}_l^m$. Subsequently, the STAR-RIS agent collects state information from the UAV agents and determines its optimal configuration through Algorithm~\ref{alg:star_ris_sa}. Following this sequential decision process, the environment generates rewards for all agents.

\par For operational deployment, our HMCD framework exhibits decentralized execution capabilities as the agents make decisions independently. In practical implementations, the STAR-RIS can be equipped with an intelligent controller that monitors real-time UAV positional data and dynamically reconfigures transmission and reflection matrices accordingly. 

\subsection{Complexity Analysis of HMCD}

\par In this section, we analyze the computational and space complexity of HMCD algorithm during both training and execution phases, following the approach in~\cite{Xie2025}.

\par The computational complexity of HMCD during the training phase is $\mathcal{O}(N_EN_LN_UN_M(N_Md(|\boldsymbol{s}| + |\boldsymbol{a}|) + {N_M}^2d + N_Md(2d) + |\boldsymbol{\theta_m}| + |\boldsymbol{\bar \theta_m}| + |\boldsymbol{\psi_m}| + |\boldsymbol{\bar \psi_m}|) + N_EN_LN_SN_C)$, which can be detailed as follows:

\begin{itemize}

\item {Network Initialization:} This requires $\mathcal{O}(N_M(|\boldsymbol{\theta_m}| + |\boldsymbol{\bar \theta_m}| + |\boldsymbol{\psi_m}| + |\boldsymbol{\bar \psi_m}|))$ operations to initialize all parameters for $N_M$ UAV agents, where $|\cdot|$ indicates the parameter count in each network.

\item {Action Sampling and Noise Application:} This process has complexity $\mathcal{O}(N_EN_LN_M)$, which accounts for policy sampling and velocity guidance mechanism application across all episodes and steps.

\item {STAR-RIS Optimization:} The ATSO algorithm contributes $\mathcal{O}(N_EN_LN_SN_C)$ complexity, where $N_S$ denotes the number of STAR-RIS elements and $N_C$ represents the average number of candidates evaluated per element.

\item {Environment Interaction:} Reward calculation and state transitions requires $\mathcal{O}(N_EN_LN_MV)$ operations, with $V$ representing environment interaction complexity.

\item {Network Updates:} This phase involves multiple components. The critic network with attention mechanism requires $\mathcal{O}(N_Md(|\boldsymbol{s}| + |\boldsymbol{a}|))$ for state and action encoding, $\mathcal{O}({N_M}^2d)$ for attention computation, and $\mathcal{O}(N_Md(2d))$ for Q value calculation. Actor network updates require $\mathcal{O}(|\boldsymbol{\theta_m}|)$ operations, while target network updates need $\mathcal{O}(|\boldsymbol{\bar \theta_m}| + |\boldsymbol{\bar \psi_m}|)$ operations. These updates occur $N_U$ times per step, thereby resulting in a total update complexity of $\mathcal{O}(N_EN_LN_UN_M(N_Md(|\boldsymbol{s}| + |\boldsymbol{a}|) + {N_M}^2d + N_Md(2d) + |\boldsymbol{\theta_m}| + |\boldsymbol{\bar \theta_m}| + |\boldsymbol{\psi_m}| + |\boldsymbol{\bar \psi_m}|))$.

\end{itemize}

\par The space complexity during training is $\mathcal{O}(N_M(|\boldsymbol{\theta_m}| + |\boldsymbol{\bar \theta_m}| + |\boldsymbol{\psi_m}| + |\boldsymbol{\bar \psi_m}|) + |\boldsymbol{\mathcal{B}}|(2|\boldsymbol{s}| + |\boldsymbol{a}| + 1) + 3N_R)$. This accounts for neural network parameters, replay buffer storage, and STAR-RIS configuration. For each STAR-RIS element, we store three parameters, two phase shifts $\theta^{R}$ and $\theta^{T}$, plus one amplitude parameter either $a^{R}$ or $a^{T}$ since the other can be calculated using $a^{R} + a^{T} = 1$. The replay buffer stores experience tuples containing states, actions, rewards, and next states.

\par During execution phase, the computational complexity reduces to $\mathcal{O}(N_M + N_RN_C)$, covering action selection from trained policies and STAR-RIS optimization. The corresponding space complexity is $\mathcal{O}(N_M|\boldsymbol{\theta_m}| + 3N_R)$, which accounts for actor network parameters and the three parameters per STAR-RIS element. This efficient execution profile enables practical deployment in time sensitive aerial communication applications.

%
\section{Simulation Results and Analysis}\label{sec:simulation}
 
\par In this section, we evaluate the performance of the HMCD and compare it with the baselines. Specifically, we first demonstrate the convergence properties of HMCD. Subsequently, we present a comparative performance evaluation against state-of-the-art baselines to establish the relative efficacy of our HMCD. Finally, we present some visualization results.

\subsection{Simulation Setups}

\begin{figure}[htbp] 
	\centering
	\subfloat[]{\includegraphics[width=1.0\linewidth]{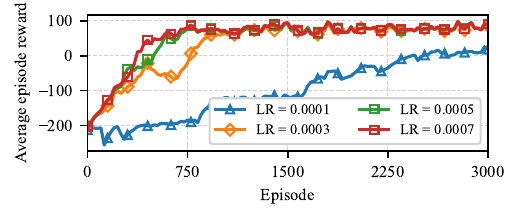}}
	\label{fig:convergence_lr}\\  
	\subfloat[]{\includegraphics[width=1.0\linewidth]{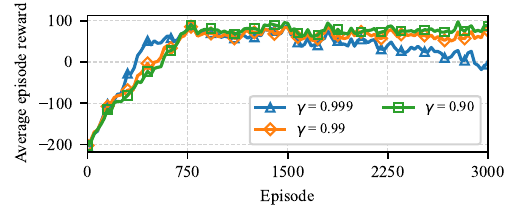}}
	\label{fig:convergence_gamma}\\  
	\subfloat[]{\includegraphics[width=1.0\linewidth]{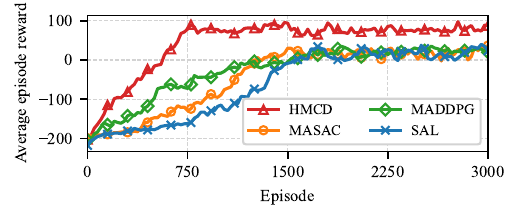}}
	\label{fig:convergence}
	
	\caption{Hyperparameter tuning results and convergence performance obtained by different methods. (a) Convergence verification versus different learning rate. (b) Convergence verification versus different discount factors. (c) Convergence performance comparison.}
	\label{Fig:convergences}
\end{figure}

\par We evaluate HMCD in a simulated low-altitude urban environment, which consists of UAV swarm, a STAR-RIS mounted on a building rooftop, and mobile users distributed on both sides of the STAR-RIS. Specifically, the simulation parameters are organized into environmental and algorithmic configurations as follows.


\par In terms of environmental parameters, the UAV operational area spans 100 m × 100 m with altitude 60-90 m. STAR-RIS is located at coordinates (1500 m, 1500 m, 20 m). Users follow GMRMM model with same-side movement range [1480 m, 1530 m] × [1400 m, 1490 m] and opposite-side range [1480 m, 1530 m] × [1510 m, 1600 m]. The UAV swarm consists of 8 vehicles with 0.5 m minimum separation. STAR-RIS contains 60 reconfigurable elements. The communication system operates at 2.4 GHz with 0.1 W transmit power, 2 MHz bandwidth, and -155 dBm/Hz noise power spectral density. Path loss parameters are 2.7 for reflection and 3.6 for direct paths. UAV energy parameters comprise 2 kg mass, 120 m/s blade tip speed, 4.03 m/s rotor induced velocity, 1.225 kg/m$^3$ air density, 0.503 m$^3$ rotor disc area, 0.6 fuselage drag ratio, and 0.05 rotor solidity.

\par In relation to algorithm parameters, for the HMCD algorithm, we configure the SA process with initial temperature 1.0, cooling rate 0.95, and minimum temperature 0.1. Additionally, learning parameters include episode count $N_E = 3000$, target network update coefficient $\tau = 0.005$, temperature parameter $\alpha = 0.01$, batch size $N_B = 256$, discount factor $\gamma = 0.90$, and learning rate 0.0007.

\par For comparative evaluation, we implement several baseline approaches using identical parameter settings. All baseline methods incorporate our proposed STAR-RIS control policy ATSO to ensure a fair comparison focused specifically on the multi-agent coordination aspects of the HMCD.

\begin{itemize}
    \item \textit{MADDPG:}  A classical MADRL approach employing the centralized training with decentralized execution (CTDE) paradigm~\cite{pmlr-v80-haarnoja18bMADDPG}. For exploration, MADDPG utilizes Gaussian noise with a standard deviation of 0.01 applied to the deterministic policy outputs.
    
    \item \textit{MASAC:} This approach implements the CTDE architecture similar to MADDPG but incorporates entropy regularization and stochastic policies as detailed in Section~\ref{subsubsection:MASAC}. MASAC maintains separate critic networks for each agent while sharing environmental information during training.
    
    \item \textit{Single-Agent Learning (SAL):} To demonstrate the effectiveness of multi-agent coordination, we implement a comparative approach where each UAV operates an independent SAC algorithm~\cite{SAC2}. Different from MASAC, SAL lacks the centralized training mechanism, with each agent optimizing its policy based solely on individual observations without considering the policies of other agents.
    
    \item \textit{Random}: A non-learning baseline where each agent selects random actions from the action space at each decision timestep, which provides a performance floor for evaluation purposes.
\end{itemize}

%
\subsection{Hyperparameter Tuning}

\par First, we investigate the impact of the learning rate on the performance of the HMCD. As illustrated in Fig.~\ref{Fig:convergences}(a), a higher learning rate accelerates convergence but may cause the policy to fall into a local optimum, whereas a lower learning rate ensures stable convergence at the expense of prolonged training time. Experimental results indicate that a learning rate of 0.0007 achieves a good trade-off, which offers faster convergence while maintaining training stability. Therefore, this value is adopted in subsequent simulations to balance training efficiency and performance.
\par Furthermore, Fig.~\ref{Fig:convergences}(b) illustrates the impact of the discount factor on the performance of the HMCD. A larger discount factor (\textit{e.g.}, $\gamma = 0.999$) results in faster initial convergence but leads to increased fluctuations in later stages, thereby affecting stability. In contrast, a smaller discount factor (\textit{e.g.}, $\gamma = 0.90$) yields a more stable training process. Considering both convergence stability and long-term rewards, we set the discount factor to 0.90 to achieve a balanced performance.

\subsection{Simulation Results}

\begin{figure}
\centering
\begin{minipage}[t]{1\linewidth}
  \centering
  \setlength{\abovecaptionskip}{-5pt}
  \includegraphics[width=3in]{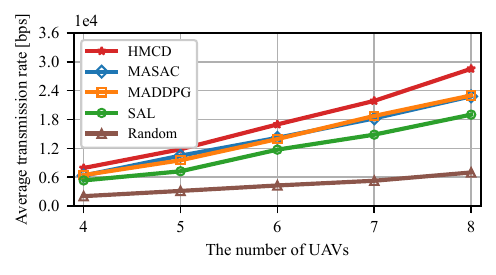}
  \caption{Average transmission rate under different number of UAVs.}
  \label{fig:obj1}
\end{minipage}
\begin{minipage}[t]{1\linewidth}
  \centering
  \setlength{\abovecaptionskip}{-5pt}
  \includegraphics[width=3in]{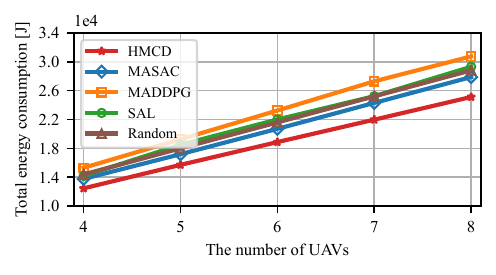}
  \caption{Total energy consumption under different number of UAVs.}
  \label{fig:obj3}
\end{minipage}
\end{figure}

\par To assess the effectiveness of our HMCD framework in solving the JREOP, we conduct comparative experiments against benchmark algorithms under identical environmental and parameter settings. This evaluation measures key performance metrics to quantify the relative advantages of our proposed HMCD across various operational conditions.

\subsubsection{Convergence Performance Analysis}

\par Fig.~\ref{Fig:convergences}(c) demonstrates the convergence characteristics of all algorithms during training. Notably, HMCD significantly outperforms the baselines, which achieves 200\% higher rewards than MADDPG and 125\% higher than MASAC, while converging 40\% faster than MADDPG and 50\% faster than MASAC. 
\par The performance differences reveal critical insights about algorithm capabilities. Specifically, MADDPG exhibits stable yet slow convergence with mid-stage fluctuations due to exploration inefficiencies in high-dimensional spaces, whereas MASAC, although faster than MADDPG, still produces suboptimal policies due to limited exploration effectiveness. Consequently, all CTDE-based methods outperform SAL, which validates our joint state-action evaluation approach. 

\subsubsection{Impact of UAV Quantity on System Performance} 

\par After establishing convergence properties, we evaluated the performance of HMCD, MASAC, MADDPG, SAL, and Random algorithms under varying numbers of UAVs, with a focus on system transmission rate and UAV energy efficiency metrics.

\par For transmission rate (Fig.~\ref{fig:obj1}), all algorithms show proportional improvements with increasing UAV quantity, but HMCD maintains a consistent advantage that amplifies as UAV numbers grow. Meanwhile, HMCD consistently outperforms all competitors, with its advantage amplifying as UAV numbers increase. Additionally, MADRL-based approaches generally surpass SAL and Random methods, which validates the efficacy of MADRL-based approaches in resource allocation tasks. Notably, despite architectural differences, MASAC and MADDPG yield comparable transmission improvements, likely due to their shared CTDE framework and similar exploration constraints in high-dimensional state spaces.

\par For energy efficiency (Fig.~\ref{fig:obj3}), all algorithms exhibit approximately linear energy consumption growth with more UAVs, which aligns with our theoretical energy consumption model. Specifically, HMCD achieves optimal energy efficiency through its sophisticated optimization mechanism, which includes precision energy allocation via the self-attention evaluation and reduced ineffective movements through the adaptive velocity transition mechanism. Furthermore, there are two critical observations to note. First, MADDPG exhibits unexpectedly poor energy optimization, which performs worse than even Random allocation, potentially due to computational overhead from frequent policy updates in continuous action space. Second, SAL only marginally outperforms Random allocation, thereby suggesting its local optimization approach fundamentally limits global energy conservation potential.

\subsubsection{Impact of STAR-RIS Element Quantity on System Performance}

\par
\begin{figure}
\begin{minipage}[t]{1\linewidth}
  \centering
  \setlength{\abovecaptionskip}{-5pt}
  \includegraphics[width=3in]{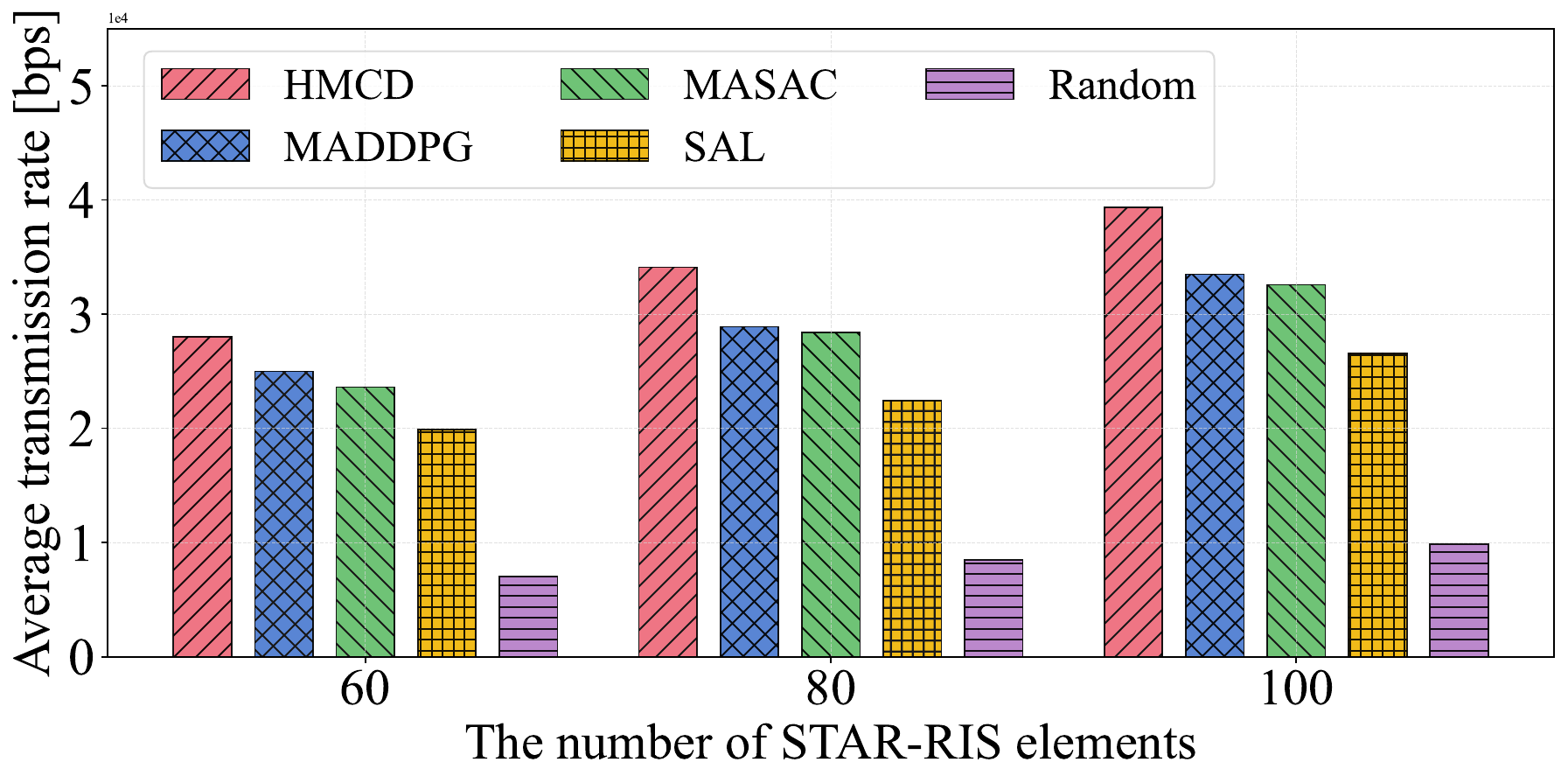}
  \caption{Average transmission rate under different number of STAR-RIS elements.}
  \label{fig:star-ris_obj1}
\end{minipage}
\end{figure}

 In addition to the number of UAVs, the number of STAR-RIS elements is also a key factor affecting the performance of the algorithms. As shown in Fig.~\ref{fig:star-ris_obj1}, there is a positive correlation between the quantity of STAR-RIS elements and the algorithm performance. 

\par To be more specific, when elements increase from 60 to 100, the transmission rate of HMCD improves by 39.3\%, compared to 37.5\%, 34.7\%, and 30.0\% for MASAC, MADDPG, and SAL respectively. Apart from that, while Random allocation shows a 66.7\% improvement rate, its absolute performance remains significantly lower. As a result, HMCD ranks first in both absolute and relative increments, which indicates that increasing elements is beneficial to improving performance, but a suitable intelligent optimization algorithm is required to give full play to its potential. 

\begin{figure}
  \centering
  \includegraphics[width=3.3in]{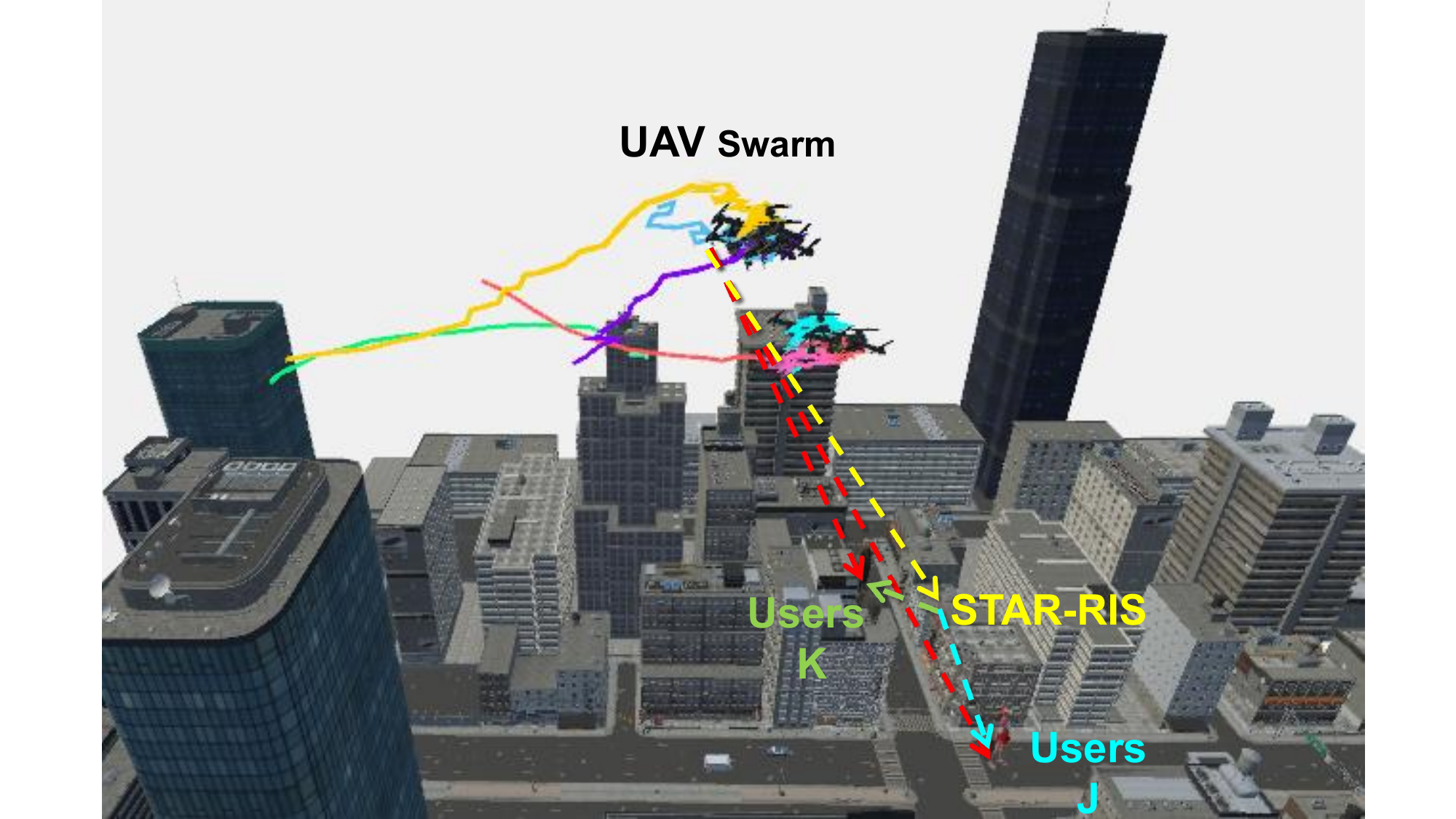}
  \caption{The trajectories of UAVs obtained by HMCD with Unity 3D. As can be seen, the UAVs tend to fly close to the STAR-RIS.}
  \label{fig:tra}
\end{figure}

\subsubsection{3D Visualization of UAV Flight Trajectories}

\par To further validate the efficacy of HMCD in wireless communication networks, we implement a high-fidelity simulation, which uses Unity 3D to model complex urban environments and render UAV flight trajectories. 

\par Fig.~\ref{fig:tra} shows the optimized flight trajectories of the UAV swarm under HMCD control. The visualization reveals a clear pattern wherein the UAV swarm systematically approaches the STAR-RIS as closely as possible while maintaining strict adherence to designated flight area boundaries. This behavior, in turn, provides compelling visual confirmation that the directional guidance terms we proposed effectively guide exploration strategies. Taken together, these observations validate the capability of HMCD to execute efficient and safe communication path planning, even in complex urban environments where direct line-of-sight paths may be obstructed by buildings and other infrastructure.

%
\section{Conclusion}\label{sec:conclusion}

\par In this paper, we have investigated the challenge of obstacle-induced link vulnerabilities in urban UAV communication networks by integrating UAV swarm with STAR-RIS technology. We have first formulated the problem as a JREOP, which aims to maximize the transmission rate of the overall system while minimizing the energy consumption of the UAV swarm. To address this, we have proposed the HMCD framework, which synergistically combines ATSO for STAR-RIS optimization with enhanced-MASAC for coordinated UAV control, further strengthened by self-attention evaluation and an adaptive velocity transition mechanism. Extensive simulation results have demonstrated that HMCD consistently outperforms baseline methods in terms of convergence speed, transmission efficiency, and energy savings. Furthermore, our analysis has indicated that system performance improves with increasing numbers of UAVs and STAR-RIS elements. Finally, through visualizing the HMCD-optimized UVAA flight trajectories in Unity 3D, we have verified the capability of HMCD in guiding UVAA to approach STAR-RIS while maximizing distance from boundaries. Looking ahead, future work will focus on enhancing anti-eavesdropping capabilities, exploring multi-STAR-RIS deployments, and incorporating advanced generative artificial intelligence models to tackle high-dimensional resource optimization in next-generation aerial communication networks.
\bibliographystyle{IEEEtran}
\bibliography{main}
\end{document}